\documentclass[a4paper,twoside]{jpconf}

\usepackage[english]{babel}
\usepackage[latin1]{inputenc}
\usepackage{times}
\usepackage[T1]{fontenc}
\usepackage{amsmath,amssymb,xspace}
\usepackage{pstricks,pst-plot}
\usepackage{graphics}
\usepackage{psfrag}
\usepackage{fixmath}
\usepackage{bbm}
\usepackage{dsfont}
\usepackage{mathrsfs}
\usepackage{color}
\usepackage{graphicx}
\usepackage{amsthm}

\theoremstyle{definition}
\newtheorem{Def}{Definition}[section]

\theoremstyle{plain}

\newtheorem{Thm}[Def]{Theorem}

\theoremstyle{remark}

\newtheorem*{Thm*}{}

\newcommand{\rW}{\rho}
\newcommand{\zW}{\zeta}
\newcommand{\pmN}{^\pm_\mathrm{N}}
\newcommand{\pN}{^+_\mathrm{N}}

\newcommand{\pS}{^+_\mathrm{S}}

\newcommand{\beq}{\begin{equation}}
\newcommand{\eeq}{\end{equation}}
\newcommand{\bea}{\begin{eqnarray}}
\newcommand{\eea}{\end{eqnarray}}
\newcommand{\bit}{\begin{itemize}}
\newcommand{\eit}{\end{itemize}}
\newcommand{\ben}{\begin{enumerate}}
\newcommand{\een}{\end{enumerate}}

\newcommand{\rh}{r_\mathrm{h}}
\newcommand{\dd}{\mathrm{d}}
\newcommand{\ee}{\mathrm{e}}
\newcommand{\ii}{\mathrm{i}}

\newcommand{\Hp}{\mathcal H^+}
\newcommand{\Hm}{\mathcal H^-}
\newcommand{\Hpm}{\mathcal H^\pm}

\newcommand{\Ap}{A^+}
\newcommand{\Am}{A^-}

\usepackage{fancyhdr} 
\begin{document}
\title{The interior of axisymmetric and stationary black holes: Numerical and analytical studies}

\author{Marcus Ansorg}

\address{Theoretisch-Physikalisches
Institut, Friedrich-Schiller-Universit\"at Jena, Max-Wien-Platz 1, D-07743 Jena, Germany}

\ead{Marcus.Ansorg@uni-jena.de}

\author{J\"org Hennig}

\address{Dept.~of Mathematics and Statistics, University of Otago, P.O. Box 56, Dunedin 9054, New Zealand}

\ead{jhennig@maths.otago.ac.nz}

\begin{abstract}
We investigate the interior hyperbolic region of axisymmetric and stationary black holes surrounded by a matter distribution. First, we treat the corresponding initial value problem of the hyperbolic Einstein equations numerically in terms of a single-domain fully pseudo-spectral scheme. Thereafter, a rigorous mathematical approach is given, in which soliton methods are utilized to derive an explicit relation between the event horizon and an inner Cauchy horizon. This horizon arises as the boundary of the future domain of dependence of the event horizon. Our numerical studies provide strong evidence for the validity of the 
universal relation $\Ap\Am = (8\pi J)^2$ where $\Ap$ and $\Am$ are the areas of event and inner Cauchy 
horizon respectively, and $J$ denotes the angular momentum. With our 
analytical considerations we are able to prove this relation rigorously.
\end{abstract}

\section{Introduction}
Axisymmetric and stationary black hole space-times are characterized by the existence of two Killing vectors
$\xi$ and $\eta$. Outside the black hole, these vectors can always be linearly combined to
form a time-like vector. In contrast, any such non-trivial linear combination, taken in some {\em interior} 
neighborhood of the event horizon $\Hp$, inevitably leads to a space-like vector. The corresponding Einstein equations, expressed in Boyer-Lindquist-type coordinates, are elliptic in the black hole's exterior and hyperbolic in its interior. It has been shown in \cite{Ansorg_Hennig2008} that for non-vanishing angular momentum $J$ of the black hole, there always exists a regular boundary of the future domain of dependence of the event horizon, the `inner Cauchy horizon' $\Hm$.

In this contribution, we investigate the interior region of axisymmetric and stationary black holes which are surrounded by a matter distribution. The  Einstein field equations are considered in Boyer-Lindquist-type, as well as Weyl coordinates, where they possess a well-defined behavior at the horizons. Note that in Boyer-Lindquist-type coordinates, the mathematical form of the field equations is the same at $\Hp$ and $\Hm$.

In section \ref{sec:numerics} we apply a fully pseudo-spectral method and solve the degenerate hyperbolic Einstein equations as an initial value problem. We expand the field quantities with respect to both spatial and time coordinates in terms of Chebyshev polynomials. Starting at the past Cauchy horizon $\Hp$, we evolve the data into the black hole interior up to the future Cauchy horizon $\Hm$. The data, found in this manner at $\Hm$, obey the remarkable property
\beq\label{ApAm} \Ap\Am = (8\pi J)^2 \,,\eeq
 where $\Ap$ and $\Am$ are the areas of event and inner Cauchy horizon respectively. The numerical solutions possess the extreme accuracy known for spectral methods when applied to elliptic problems. In particular, the relation (\ref{ApAm}) was found to be valid up to 12 decimal digits.
 
In section \ref{sec:analytics} we review analytical work through which (\ref{ApAm}) was proved rigorously \cite{Ansorg_Hennig2008}. The corresponding techniques are based on `B\"acklund transformations', a method known from soliton theory. We conclude this article with a discussion of generalized and related black hole properties in axisymmetry.

\section{Numerical studies}\label{sec:numerics}

In Boyer-Lindquist-type coordinates $(R,\theta,\varphi,t)$, the metric becomes singular at $R=\pm\rh$,
	\beq\label{BoyerL}
		\dd s^2 = \hat{\mu}\left(\frac{\dd R^2}{R^2-\rh^2}+\dd\theta^2\right)+\hat{u}\sin^2\theta(\dd\varphi^2-\omega \dd t)^2-\frac{4}{\hat{u}}(R^2-\rh^2)\dd t^2\,,
	\eeq	
 where $\pm\rh$ denote the coordinate locations of the two horizons $\Hpm$. 
The interior hyperbolic region of the black hole is characterized by 
	\beq R\in(-\rh,\rh),\qquad\theta\in[0,\pi]\,.\eeq
Here, the metric coefficients $\hat{\mu}, \omega, \hat{u}$ depend on $R$ and $\theta$ only and are regular at $R=\pm\rh$ (see \cite{Carter, Bardeen}). The corresponding hyperbolic Einstein equations read as follows: 
\bea
\label{EG1} (R^2-\rh^2)\tilde u_{,RR}+2R\tilde u_{,R}+\tilde u_{,\theta\theta}
 +\tilde u_{,\theta}\cot\theta 
  &=& 1 - \frac{\hat{u}^2}{8}\sin^2\!\theta\!
  \left(\omega_{,R}^2+\frac{\omega_{,\theta}^2}{R^2-\rh^2}\right)\,, \\[4mm]
\label{EG2}  (R^2-\rh^2)(\omega_{,RR}+4\omega_{,R}\tilde u_{,R})
 +\omega_{,\theta\theta}&=& -\,\omega_{,\theta}(3\cot\theta+4\tilde u_{,\theta})\,,\\[4mm]
\label{EG3} (R^2-\rh^2)\tilde\mu_{,RR}+R\tilde\mu_{,R}
 +\tilde\mu_{,\theta\theta}
 & = & \frac{\hat{u}^2}{16}\sin^2\!\theta
  \left(\omega_{,R}^2+\frac{\omega_{,\theta}^2}{R^2-\rh^2}\right)
   + R\tilde u_{,R}\nonumber\\
  & &  -(R^2-\rh^2)\tilde u_{,R}^2
  -\tilde u_{,\theta}(\tilde u_{,\theta}+\cot\theta),
\eea
where $\tilde u:=\frac{1}{2}\ln\left(\hat u/\hat u\pN\right)$ and
$\tilde\mu:=\frac{1}{2}\ln\left(\hat\mu/\hat u\pN\right)$ with $\hat u\pN:=\hat{u}(R=\rh,\theta=0)$\footnote{Here we choose the north pole value $\hat u\pN$ of $\hat u$ as normalization constant. It remains positive and finite even in the degenerate limit $\rh\to 0$.}. 
Note that $\hat{\mu}$ does not enter in the equations (\ref{EG1}, \ref{EG2}). An explicit solution of this system is given by the Kerr metric which describes a single rotating black hole in vacuum.

At $R=\pm\rh$, the metric potentials have to obey the following boundary conditions (see \cite{Carter, Bardeen})
\beq\label{BC}
 \Hpm:\quad\left\{\begin{array}{l}  
				\omega=\textrm{constant}=\omega^\pm\,,\\[4mm] 
						\pm\,2\,\rh\,\tilde{u}_{,R}\,+\,\tilde{u}_{,\theta\theta}\,+\,\tilde{u}_{,\theta}\cot\theta
							=1-\frac{1}{8}\hat{u}^2\omega_{,R}^2\sin^2\theta
				\end{array}\right.
\eeq
with the horizon angular velocities $\omega^\pm$. 

Expressions for the black hole angular momentum $J$ and horizon areas $A^\pm$ have been discussed in \cite{Carter,Bardeen}. They are given by 
\bea\label{J}
  \fl J  &=&  \frac{1}{8\pi}\oint\limits_\mathcal{H^\pm}\eta^{a;b}\dd S_{ab}
     = -\frac{1}{16}\int\limits_0^\pi\hat u^2\omega_{,R}\big|_\mathcal{H^\pm}
       \sin^3\!\theta\,\dd\theta,\\
  \fl\label{A} A^\pm &=& 2\pi\int\limits_0^\pi\sqrt{\hat\mu\hat u}\big|_\mathcal{H^\pm}
     \sin\theta\,\dd\theta
   = 4\pi \hat u\pmN,
\eea
where the `north pole' values of $\hat{u}$ are denoted by \[\hat u\pmN=\hat{u}(R=\pm\rh,\theta=0).\]
Note that, for the area formula (\ref{A}), we used the following relations:
\beq \begin{array}{llcll}
				 \mbox{At the Horizons $\Hpm\,(R=\pm\rh)$}&:& \sqrt{\hat\mu\hat u} &=& \textrm{constant} \\[4mm]
				 \mbox{At the Rotation axis $(\sin\theta=0)$}&:& \hat\mu&=&\hat u\,.
		\end{array}
\eeq
In our numerical scheme we solve the equations (\ref{EG1}) and (\ref{EG2}). 
The corresponding solutions provide us with the relevant information to compute $J$ and $A^\pm$. 

\begin{figure}
		\centering
		\psfrag{Time direction}[tc][c]{{`Time' direction}}
		\psfrag{R}[b][r]{{$R$}}
		\psfrag{th}[r][t]{{$\theta$}}
		\psfrag{pi}[r][t]{{$\pi$}}
		\psfrag{0}[r][t]{{$0$}}
		\psfrag{-rh}[rt][b]{{-$\rh$}}
		\psfrag{+rh}[ct][b]{{$\rh$}}
		\includegraphics[width=9cm, height=7cm]{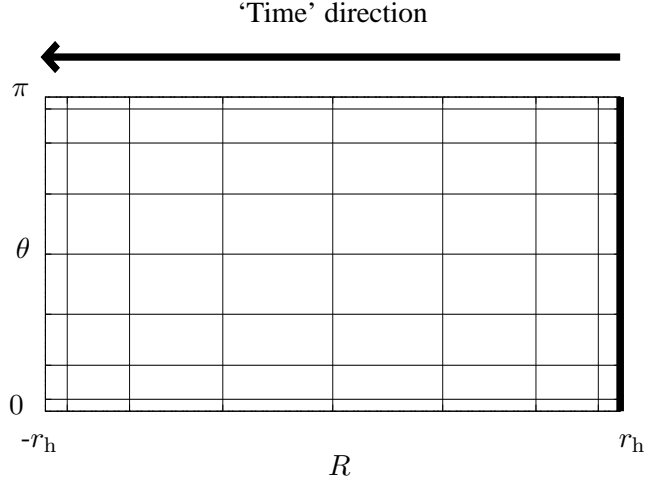}
		\caption{\label{Fig1} The interior hyperbolic black hole region with past $(R=\rh)$ and future ($R=-\rh$) Cauchy horizons ${\Hpm}$.
The metric is determined as the solution to an initial value problem starting at $R=\rh$ and evolving up 
to $R=-\rh$. We indicate the location of the Gauss-Lobatto gridpoints on the spectral grid which is used in the pseudo-spectral collocation point method.}
	\end{figure}

As a first step, we express the metric coefficients $\tilde{u}$ and $\omega$ in terms of auxiliary potentials $U$ and $\Omega$:
		\bea
			\nonumber\tilde{u}(R,\theta) &=& \tilde{u}^+(\theta)+(R-\rh)U(R,\theta) \\
			\nonumber\omega(R,\theta)    &=& \omega^+ + (R-\rh)\omega^+_R(\theta)+(R-\rh)^2\Omega(R,\theta) \,.
		\eea
In this {\em ansatz}, we take into account the boundary condition $\omega(R=\rh,\theta)=\omega^+$. We prescribe freely a set of `initial' data $\{\tilde{u}^+(\theta), \omega^+, \omega^+_R(\theta)\}$ which is then evolved from $R=\rh$ into the black hole interior up 
to $R=-\rh$ via the equations (\ref{EG1}) and (\ref{EG2}) (see figure~\ref{Fig1}). Note that the second boundary condition in (\ref{BC}) fixes the first `time' derivative $\tilde{u}_{,R}(\rh,\theta)$ in terms of $\tilde{u}^+(\theta)$ and $\omega^+_R(\theta)$, and hence $\tilde{u}_{,R}(\rh,\theta)$ cannot be prescribed freely.

We have chosen a {\em fully} pseudo-spectral scheme \cite{Hennig_Ansorg2009a} for the numerical integration of (\ref{EG1}) and (\ref{EG2}). In practice, we expand the functions $U$ and $\Omega$ with respect to Chebyshev polynomials:
		  \bea U &\approx& \sum_{j=0}^n\sum_{k=0}^n c_{jk}^{(U)} T_j\left(\frac{R}{\rh}\right) T_k\left(\frac{2}{\pi}\,\theta-1\right) \\[4mm]
		     \Omega &\approx& \sum_{j=0}^n\sum_{k=0}^n c_{jk}^{(\Omega)} T_j\left(\frac{R}{\rh}\right) T_k\left(\frac{2}{\pi}\,\theta-1\right) \eea
 and consider the field equations on an $n\times n$-`spectral grid' with Gauss-Lobatto gridpoints $(R_j,\theta_k)$ (see figure~\ref{Fig1}),
\beq  
		R_j = \rh\cos\frac{\pi j}{n-1}\,,\qquad \theta_k=\pi\sin^2\frac{\pi k}{2(n-1)}\qquad(j,k=0,\ldots,n-1)\,.
\eeq
The pseudo-spectral collocation point method provides an approximation of the values 
\beq
	U_{jk} = U\left(R_j,\theta_k\right)\,,\quad \Omega_{jk} = \Omega\left(R_j,\theta_k\right)
\eeq
which we collect in the $2n^2$-dimensional vector
\beq
	{\bf f} = \left(\begin{array}{c} 
						U_{00} \\[2mm] \Omega_{00} \\[2mm] \vdots \\[2mm] U_{NN} \\[2mm] \Omega_{NN}
					\end{array}
			\right) \,,\qquad N=n-1\,.
\eeq
From any such vector ${\bf f}$ the spectral coefficients $c_{jk}^{(U)}$ and $c_{jk}^{(\Omega)}$ of $U$ and $\Omega$ can be computed. 
Moreover, the corresponding coefficients of first and second derivatives of $U$ and $\Omega$ are easily determined. 
Returning from coefficient to physical space, we can build vectors ${\bf f}_{,R},\ldots,{\bf f}_{,\theta\theta}$ containing the values of  first and second derivatives $U_{,R},\Omega_{,R},\ldots,U_{,\theta\theta},\Omega_{,\theta\theta}$ at all collocation points $(R_j,\theta_k)$. 

	\begin{figure}[h]
		\centering
		\psfrag{ut0}[B][b]{{\footnotesize$\tilde{u}^+$}}
		\psfrag{om1}[B][b]{{\footnotesize$\rh^2\omega^+_R$}}
		\psfrag{th}[c][r]{{$\theta$}}
		\includegraphics[width=7.3cm, height=5cm]{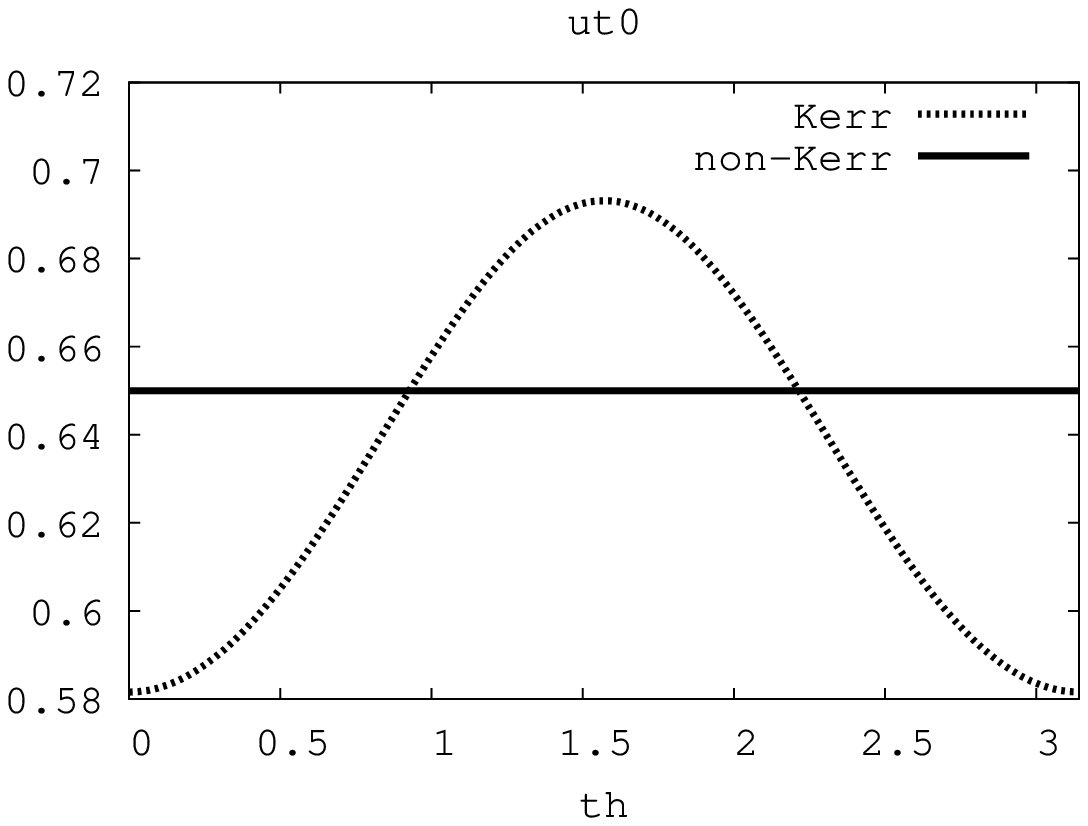}
		\includegraphics[width=7.3cm, height=5cm]{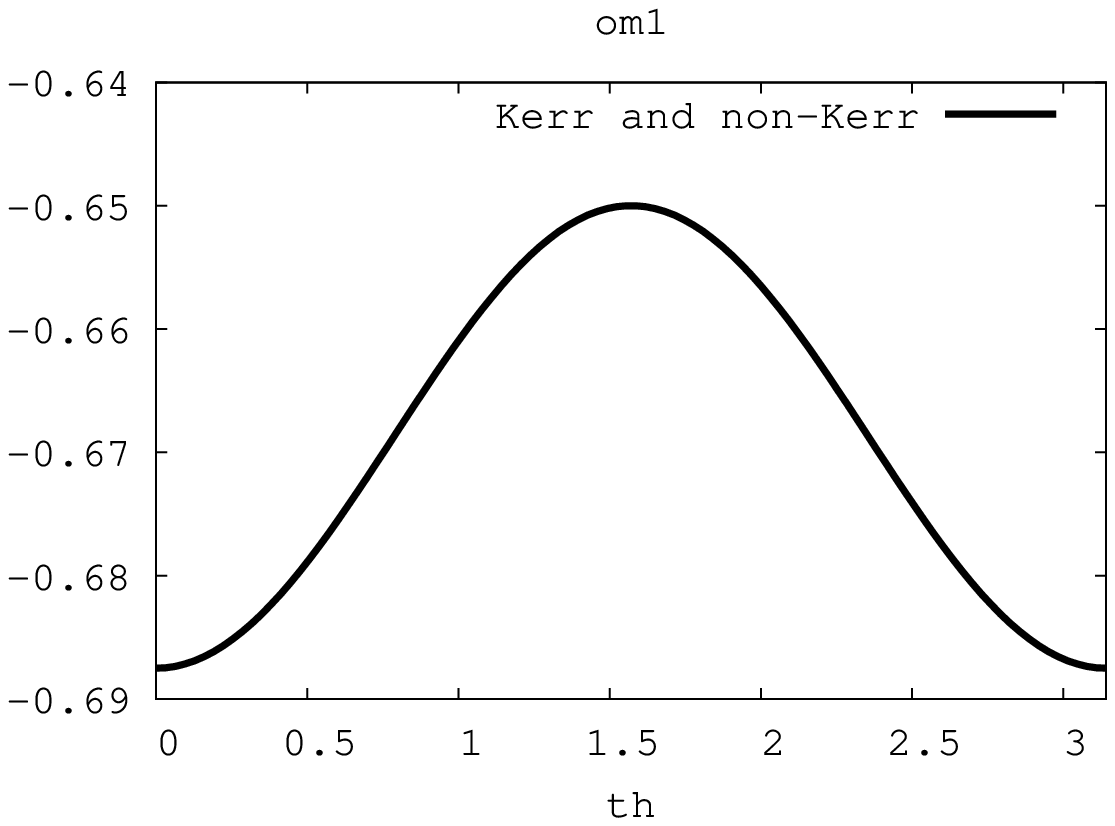}\vspace*{5mm}
%
%
		\psfrag{ut}[B][b]{{\footnotesize$\tilde{u}$}}
		\psfrag{om}[B][b]{{\footnotesize$\rh\omega$}}
		\psfrag{th}[c][r]{{$\theta$}}
		\includegraphics[width=7.3cm, height=5cm]{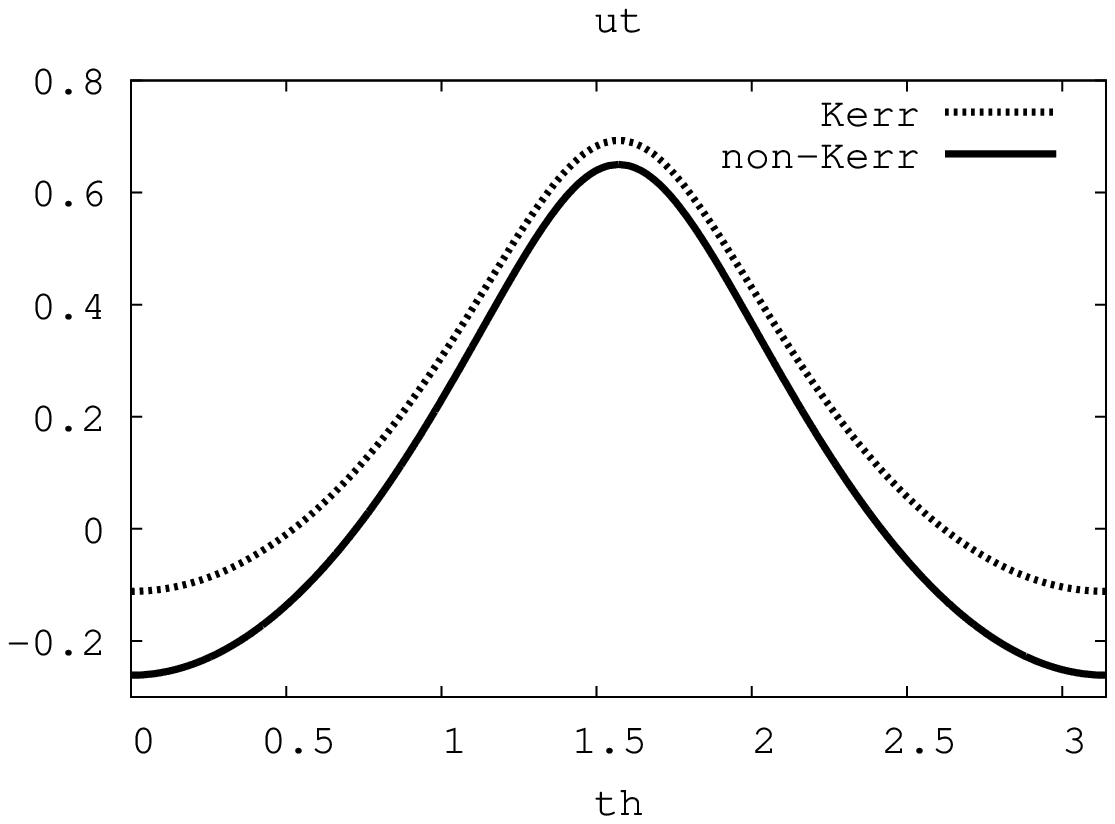}
		\includegraphics[width=7.3cm, height=5cm]{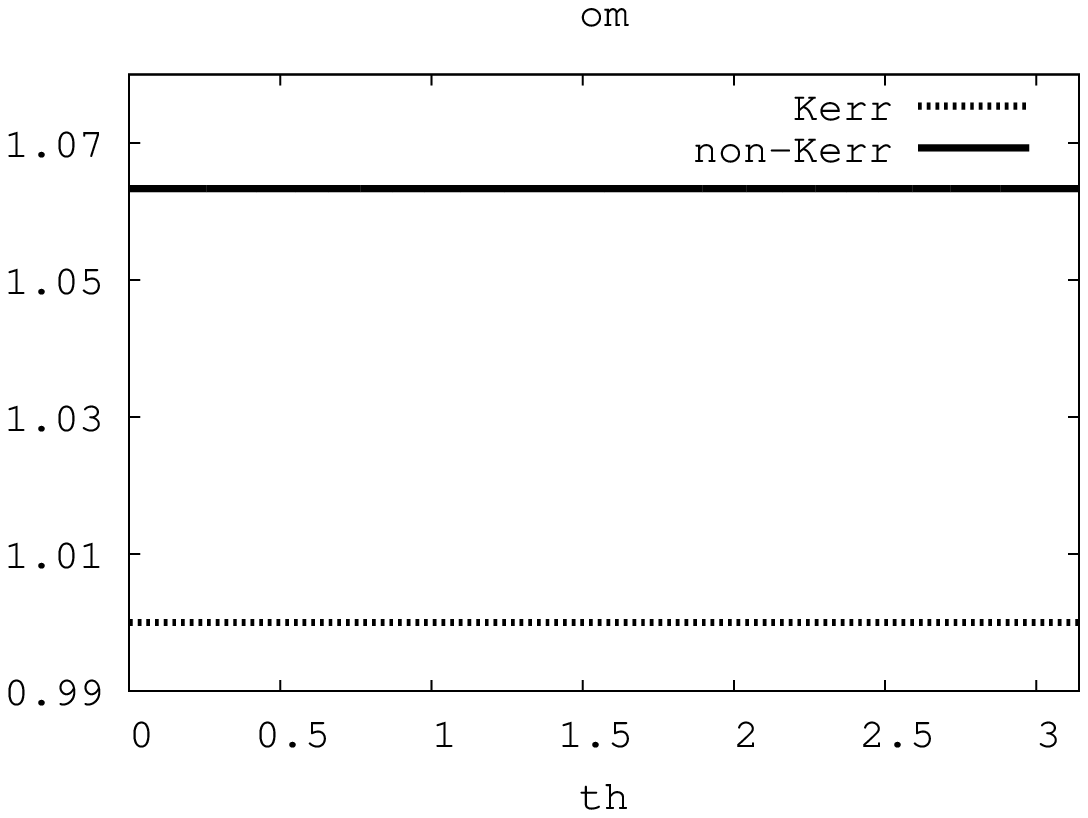}
		\caption{\label{Fig3} Initial data for Kerr and non-Kerr solution (upper panels) and corresponding values at the future Cauchy horizon $\Hm$ (lower panels).}
	\end{figure}
Now consider the system (\ref{EG1}, \ref{EG2}) at $(R_j,\theta_k)$ and insert the relevant components of ${\bf f}, {\bf f}_{,R},\ldots,{\bf f}_{,\theta\theta}$ into these equations. We obtain a non-linear algebraic system of equations of the form
\beq\label{F_of_f}
	{\bf F(f)} = 0\,,
\eeq
where 
\beq
	{\bf F} = \left(\begin{array}{c} 
						F^{(U)}_{00} \\[2mm] F^{(\Omega)}_{00} \\[2mm] \vdots \\[2mm]  F^{(U)}_{NN} \\[2mm] F^{(\Omega)}_{NN}
					\end{array}
			\right) \,.
\eeq
Here, $F^{(U)}_{jk}$ and $F^{(\Omega)}_{jk}$ denote the difference of left and right hand sides of equations (\ref{EG1}) and (\ref{EG2}) respectively, taken at the collocation point $(R_j,\theta_k)$. The solution ${\bf f}$ of the discrete algebraic system (\ref{F_of_f}) describes the spectral approximation of the solution to our hyperbolic initial value problem. 

We find the vector ${\bf f}$ using a Newton-Raphson scheme,
\beq\label{NewtRaph}
	{\bf f} = \lim_{m\to\infty}{\bf f}_m\,,\quad
	{\bf f}_{m+1} = {\bf f}_m - \left[{\bf J}({\bf f}_m)\right]^{-1} 
		{\bf F}\left({\bf f}_m\right)\,,
\eeq
where the Jacobian matrix is given by
\beq\label{Jacobian}
	{\bf J} = \frac{\partial{\bf F}}{\partial{\bf f}} \,.
\eeq

Note that for the convergence of the scheme a `good' initial guess ${\bf f}_0$ is necessary. For the first calculation, we take the corresponding coefficients of the Kerr solution for some specific parameters (say $J$ and $A^+$). We may then depart from the Kerr solution and gradually approach some new solution with a non-Kerr initial data set $\{\tilde{u}^+(\theta), \omega^+, \omega^+_R(\theta)\}$.

In the following example we start with a Kerr black hole in which we prescribe the specific angular momentum $J/M^2=0.8$, where $M$ denotes the black hole mass. As explained above, we gradually depart from this solution and approach an initial data set in which we replace the initial function $u^+(\theta)$ by a constant (we take $\tilde{u}^+(\theta)=0.65$) but 
maintain $\omega^+$ and $\omega^+_R(\theta)$ as previously determined for the Kerr case of $J/M^2=0.8$.
 The initial data $\tilde{u}^+(\theta)$ and $\omega^+_R(\theta)$, as well as the functions $\tilde{u}$ and $\omega$, at $\Hm$, are displayed in figure~\ref{Fig3}. Here, the values at the inner Cauchy horizon $R=-\rh$ emerge from the pseudo-spectral calculation.

\begin{figure}
		\centering
		\psfrag{|eq|}[B][b]{{$\left|1-(8\pi J)^2/(A^+A^-)\right|$}}
		\psfrag{n}[c][c]{{\normalsize$n$}}
		\psfrag{ 5}[c][c]{{\normalsize$5$}}
		\psfrag{ 10}[c][c]{{\normalsize$10$}}
		\psfrag{ 15}[c][c]{{\normalsize$15$}}
		\psfrag{ 20}[c][c]{{\normalsize$20$}}
		\psfrag{ 25}[c][c]{{\normalsize$25$}}
		\psfrag{ 30}[c][c]{{\normalsize$30$}}
		\psfrag{ 35}[c][c]{{\normalsize$35$}}
		\psfrag{ 0.01}[c][c]{{\small$10^{-2}$}}
		\psfrag{ 0.001}[c][c]{{}}
		\psfrag{ 0.0001}[c][c]{{\small\quad$10^{-4}$}}
		\psfrag{ 1e-05}[c][c]{{}}
		\psfrag{ 1e-06}[c][c]{{\small\,\,$10^{-6}$}}
		\psfrag{ 1e-07}[c][c]{{}}
		\psfrag{ 1e-08}[c][c]{{\small\,\,$10^{-8}$}}
		\psfrag{ 1e-09}[c][c]{{}}
		\psfrag{ 1e-10}[c][c]{{\small\,\,\,\,\,$10^{-10}$}}
		\psfrag{ 1e-11}[c][c]{{}}
		\psfrag{ 1e-12}[c][c]{{\small\,\,\,\,\,$10^{-12}$}}
		\psfrag{ 1e-13}[c][c]{{}}
		\includegraphics[scale=0.8]{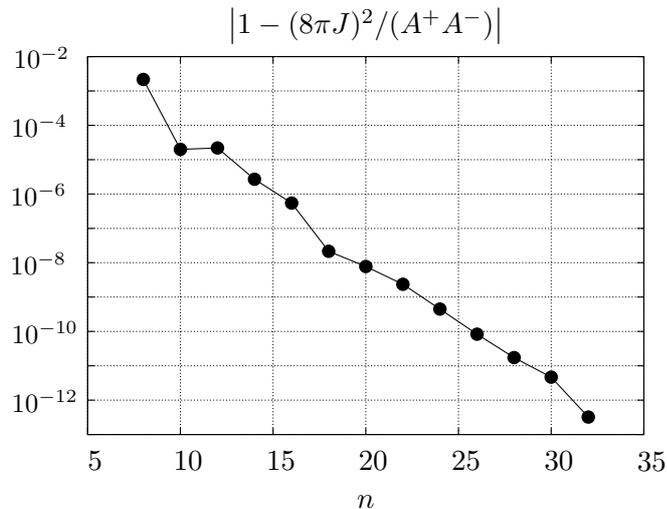}
		\caption{\label{Fig4} Numerical test of relation (\ref{ApAm}) for the non-Kerr example displayed in figure (\ref{Fig3}).}
	\end{figure}

The numerical test of relation (\ref{ApAm}) is exhibited in figure~\ref{Fig4}. It shows a convergence plot of the expression $\left|1-(8\pi J)^2/(A^+A^-)\right|$ for our sample non-Kerr solution. We obtain a confirmation of the validity of (\ref{ApAm}) by 12 relevant decimal digits which is equivalent to the numerical round-off error.

\section{Analytical studies}\label{sec:analytics}	
This section about the analytical study of the interior hyperbolic black hole region, including the rigorous derivation of the equality (\ref{ApAm}), is a brief summary of earlier work that has been presented in \cite{Ansorg_Hennig2008}.
\subsection{Weyl coordinates}
\begin{figure}
		\centering
		\includegraphics[scale=0.8]{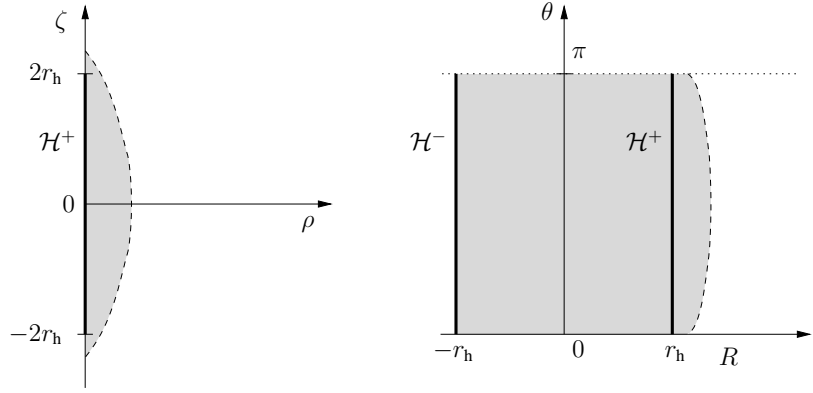}
		\caption{\label{Fig5} Portion of a black hole space-time in Weyl coordinates 
	(left panel) and Boyer-Lindquist-type coordinates
	(right panel) (figure taken from \cite{Ansorg_Hennig2008}).}
\end{figure}

As a first step towards a strict mathematical treatment in terms of {\em B\"acklund transformations}, we introduce canonical Weyl coordinates 
$(\rW,\zW,\varphi,t)$ in a small exterior vacuum vicinity of the black hole\footnote{We assume that for physically reasonable types of matter surrounding our stationary black hole, the immediate vicinity of the event
horizon must be vacuum, see. e.~g.~discussion in \cite{Bardeen}.}:
		\beq
			\rW^2 = 4(R^2-\rh^2)\sin^2\theta,\qquad
			\zW   = 2R\cos\theta.
		\eeq
The corresponding line element reads as follows:
		\beq
			\dd s^2 = \ee^{-2U}\left[\ee^{2k}(\dd\rW^2+\dd\zW^2)
						+\rW^2\dd\varphi^2\right]
					-\ee^{2U}(\dd t+a\dd\varphi)^2\,.
		\eeq
Here, the metric potentials $U, k$ and $a$ are functions of $\rW$ and  $\zW$. The rotation axis ($\sin\theta=0$) is given by:
\beq\rW=0,\qquad |\zW|\geq 2\rh\,, \eeq
and the exterior event horizon $\Hp$ is located at \beq\rW=0, -2\rh\leq\zW\leq 2\rh\,.\eeq
The event horizon is a degenerate surface in Weyl coordinates. As the interior region is characterized by $|R|<\rh$, the 
corresponding Weyl coordinate $\rW$ assumes imaginary values there. This means that the hyperbolic black hole region cannot 
be accessed in Weyl coordinates (see figure \ref{Fig5}). 
\subsection{The Ernst equation}
The complex Ernst potential $f$ combines the metric functions $U$ and $b$, 
		\beq f=\ee^{2U}+\ii b\,,\eeq
		where the twist potential $b$ is related to the coefficient $a$ via the relations:
		\beq
		a_{,\rW} = \rW\,\ee^{-4U} b_{,\zW},\qquad
		a_{,\zW} =-\rW\,\ee^{-4U} b_{,\rW}.
		\eeq
In Boyer-Lindquist-type coordinates, this relation reads as:
\beq	a_{,R} = -2\sin\theta\,\ee^{-4U}b_{,\theta},\qquad	a_{,\theta} = 2(R^2-\rh^2)\sin\theta\,\ee^{-4U}b_{,R}. \eeq
 		The vacuum Einstein equations (which are valid in a vicinity of $\Hp$) 
		are equivalent to the {\em Ernst equation} \cite {Ernst} which reads in Weyl coordinates as 
\beq	(\Re f)\left(f_{,\rW \rW} + f_{,\zW \zW} + \frac{1}{\rW}f_{,\rW}\right)		= f_{,\rW}^2 + f_{,\zW}^2 \eeq
		and in Boyer-Lindquist-type coordinates:
\beq	(\Re f)\left[(R^2-\rh^2)f_{,RR}+2Rf_{,R}+f_{,\theta\theta}+\cot\theta f_{,\theta}\right]= (R^2-\rh^2)f_{,R}^2+f_{,\theta}^2.\eeq
		Note that because of the degeneracy of $\Hp$ in Weyl coordinates, the potential $f$ is, for $\rho=0$, only a $C^0$-function in terms
		of $\zeta$. In contrast, for a regular black hole, $f$ is analytic with respect to the
		Boyer-Lindquist-type-coordinates $R$ and $\theta$.
\subsection{B\"acklund transformation}

		The Ernst equation is, when written in Weyl coordinates, the integrability condition of an associated linear matrix problem \cite{Neugebauer1979, Neugebauer1980}. This is the great advantage of the Weyl coordinates with respect to the Boyer-Lindquist-type coordinates. The existence of this linear problem enables us to apply methods known from soliton theory. In this contribution, we are particularly interested in the so-called B\"acklund transformations through which new solutions from previously known ones are created
 \cite{Harrison}--\cite{AKMN2002}. 
As an example, the Kerr solution describing a rotating black hole in vacuum can be constructed from the flat Minkowski metric in this manner (see e.~g.~\cite{Neugebauer}).

    In the following we use the B\"acklund transformation technique in order to write an arbitrary
		regular axisymmetric, stationary black hole solution $f$ in terms of an auxiliary `seed' potential $f_0$. Here, $f_0$ describes a space-time without a black hole but with a completely regular central vacuum region. More specifically, 
  		$f_0$ is characterized by the following properties:
		\begin{enumerate}
			\item $f_0$ is defined in a vicinity of the axis section
				  $\rW=0, |\zeta|\leq 2\rh$. 
			\item In this vicinity, $f_0$ is an analytic function of $\rW$
				  and $\zeta$ and an {\em even} function of $\rho$.
			\item For  $\rW=0$ and $|\zeta|\leq 2\rh$, the values of $f_0$ in terms of the event horizon values of $f$ are given by  
			\beq  \label{f0}
		      f_0=\frac{\ii\left[2\rh(b\pN+b\pS)-(b\pN-b\pS)\zW\right]f+4\rh
        b\pN b\pS} 
                 {4\rh f-\ii\left[2\rh(b\pN+b\pS)+(b\pN-b\pS)\zW\right]},
       \eeq
       where $b\pN=b(\rho=0, \zeta=2\rh)$ and $b\pS=b(\rho=0,\zeta=-2\rh)$ 
						denote the twist potential values at north and south poles of $\mathcal H^+$.
		\end{enumerate}
Now, from this Ernst potential $f_0$ the original potential $f$ can be
recovered by means of an appropriate B\"acklund transformation of the following form\footnote{A bar denotes complex conjugation.}:
\begin{equation}\label{Baecklund}
 f = \frac{\left|
   \begin{array}{ccc}
    f_0      & 1                 & 1\\
    \bar f_0 & \alpha_1\lambda_1 & \alpha_2\lambda_2\\
    f_0      & \lambda_1^2       & \lambda_2^2
   \end{array}
   \right|}
   {\left|
   \begin{array}{ccc}
    1      & 1                 & 1\\
    -1     & \alpha_1\lambda_1 & \alpha_2\lambda_2\\
    1      & \lambda_1^2       & \lambda_2^2
   \end{array}
   \right|},
\end{equation}
where
\begin{equation}\label{lambda}
 \lambda_j=\sqrt{\frac{K_j-\ii\bar z}{K_j+\ii z}},\qquad
 j=1,2,\qquad  K_1=-2\rh,\qquad  K_2=2\rh
\end{equation}
with the complex coordinates $z=\rW+\ii\zW$, $\bar z=\rW-\ii\zW$,
and $\alpha_1$, $\alpha_2$ are solutions to the Riccati equations
\bea\label{R1}
 \alpha_{j,z}&=&-(\lambda_j\alpha_j^2+\alpha_j)\frac{f_{0,z}}{f_0+\bar{f}_0}
             +(\alpha_j+\lambda_j)\frac{\bar f_{0,z}}{f_0+\bar{f}_0},\\
\label{R2}
 \alpha_{j,\bar z}&=&-\left(\frac{1}{\lambda_j}\alpha_j^2
                     +\alpha_j\right)\frac{f_{0,\bar z}}{f_0+\bar{f}_0}
             +\left(\alpha_j+\frac{1}{\lambda_j}\right)
              \frac{\bar f_{0,\bar z}}{f_0+\bar{f}_0}
\eea
with
\begin{equation}\label{norm}
 \alpha_j\bar\alpha_j=1. 
\end{equation}

For a regular black hole, $f$ is analytic with respect to $R$ and $\cos\theta$ in an exterior vicinity of ${\cal H}^+$. 
Hence, we can expand it analytically into an {\em interior} vicinity of $\mathcal H^+$. A theorem by Chru\'sciel (theorem 6.3 in \cite{Chrusciel}) assures that the potential $f$ exists as a regular solution of the interior Ernst equation for all values \footnote{We obtain Chru\'sciel's form of the line element by substituting $R=\rh\cos T$ and $\theta=\psi$.}
		\beq(R,\cos\theta)\in(-\rh,\rh]\times[-1,1]\,.\eeq
This region only excludes the Cauchy horizon $\mathcal H^-$ ($R=-\rh$). 

\begin{figure}
 \centering
 \includegraphics[scale=0.9]{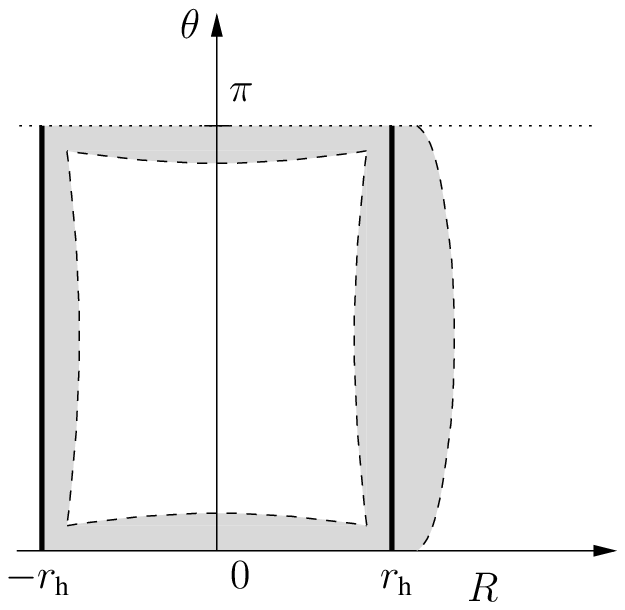}
 \caption{\label{Fig6}The seed function $f_0$ can be regularly defined
 in at least the grey
 areas (figure taken from \cite{Ansorg_Hennig2008}).}
 \label{figure2}
\end{figure}

In the following derivation of the expression for $f$ at the interior boundary $R=-\rh$, 
a crucial role is played  by the fact that $f_0$ is {\em even} in $\rho$. In terms of the Boyer-Lindquist-type coordinates, this means that $f_0$ is an analytic function of $(R^2-\rh^2)\sin^2\!\theta$ and $R\cos\theta$. The analytic expansion of $f_0$ into the region $R<\rh$ retains this property. Hence, $f_0$ taken at the boundaries of the inner hyperbolic region, can be expressed in terms of $f_0$ taken at $ R=\rh$. Specifically we obtain 
\beq
	f_0(R=\,-\,\rh\,,\cos\theta) = f_0(R=\,+\,\rh\,,-\cos\theta)=f_0(\rho=0,\zeta=-2\rh\cos\theta)\,.\,
\eeq
Also, it follows that $f_0$ is regularly defined in a sufficiently small vicinity of the boundary of
the interior region, see figure~\ref{Fig6}.

From the values of $f_0$ at these boundaries we can construct $f$ on $\mathcal H^-$ via the B\"acklund transformation, which is stated in the following.

{\Thm Any Ernst potential $f$ of a regular, axisymmetric, and stationary black hole
 space-time with angular momentum $J\neq 0$,
 can be regularly extended into the interior of the black hole up to and including an interior Cauchy horizon, described by $R=-\rh$ in Boyer-Lindquist-type coordinates $(R,\theta)$. The values of $f$ on the Cauchy horizon are given by
\begin{equation}\label{fCH}
 \fl f(R=-\rh,\cos\theta)=\frac{\ii\left[\delta_1+\delta_2
                  -(\delta_1-\delta_2)\cos\theta
                \right]f_0(R=\rh,-\cos\theta)+2\delta_1\delta_2}
                {2f_0(R=\rh,-\cos\theta)
                -\ii\left[\delta_1+\delta_2+(\delta_1-\delta_2)\cos\theta
                \right]}
\end{equation}
with
\begin{equation}
 \delta_1 = \frac{b\pS(b\pN-b\pS)+2b\pN(b_{,\theta\theta})\pN}
                 {b\pN-b\pS+2(b_{,\theta\theta})\pN},\qquad
 \delta_2 = \frac{b\pN(b\pN-b\pS)+2b\pS(b_{,\theta\theta})\pN}
                 {b\pN-b\pS+2(b_{,\theta\theta})\pN}          
\end{equation}
where the scripts {\rm `+'} and {\rm `N/S'} indicate that the corresponding value of $b$ or its second $\theta$-derivative has to be taken at the event horizon's north or south pole respectively. The values of the seed solution $f_0$ for $R=\rh$ follow via \eref{f0} from $f$ on the event horizon.\\}

Note that, for $J\to0$,  the Cauchy horizon becomes singular. In this case we have $|\delta_{1/2}|\to\infty$ and consequently $f|_{\mathcal H^-}$ diverges.

By virtue of the above theorem we are able to write the inner Cauchy horizon area $\Am$ completely in terms of Ernst expressions taken at the event horizon $\Hp$. As discussed in \cite{Ansorg_Hennig2008}, the universal equality (\ref{ApAm}) arises in this manner. Moreover, the vanishing of $\Am$ is obtained in the limit $J\to 0$, i.~e.~when the Cauchy horizon becomes singular.

\section{Discussion}

In this contribution we discussed the interior hyperbolic region of axisymmetric and stationary black holes which are surrounded by a matter distribution. We first looked at the corresponding degenerate hyperbolic Einstein equations in terms of fully pseudo-spectral methods, and confirmed the validity of relation (\ref{ApAm}) to high precision.

In the second part of the article, we used the B\"acklund technique in order to write the black hole metric in terms of an auxiliary seed potential $f_0$ that does not describe a black hole but a completely regular central region. We first derived simple relations between the values of $f_0$ at the boundaries of the interior region, and then carried these relations over to the original black hole metric by virtue of an appropriate B\"acklund transformation. A particular consequence of this relation is the universal equality (\ref{ApAm}). 

Note that the key point for this method to work is the fact that the two horizons $\Hpm$ are connected by specific axis sections. Indeed, the linear matrix problem associated to the Ernst equation (see discussion at the beginning of section 3.3) simplifies considerably for $\rho=0$, that is, on the two horizons $\Hpm$, and on the rotation axis ($\sin\theta=0$). An alternative derivation, analyzing the linear matrix problem directly at the several sections where $\rho=0$, was carried out in \cite{Ansorg_Hennig2009b,Hennig_Ansorg2009c} for the Einstein-Maxwell field. In that work, the corresponding steps yield the more general formula
\[ 
(8\pi J)^2+(4\pi Q^{\,2})^2=A^+ A^-
\] 
where $Q$ is the electric charge of the black hole.

It is interesting to remark that there are associated {\em inequalities} relating angular momentum and area of the black hole. In \cite{Hennig_etal2010} (see also \cite{Hennig_etal2008}) it was shown that for {\em subextremal} black holes (which possess trapped surfaces in every sufficiently small interior vicinity of the event horizon), the following inequalities hold:
		\[ A^- < \sqrt{(8\pi J)^2+(4\pi Q^{\,2})^2} < A^+.\]
In the case of pure Einsteinian gravity, this relation was proved in a different context, namely for non-stationary, axisymmetric, vacuum space-times \cite{DainReiris}. In particular, the local inequality $A \geq 8\pi|J|$ was shown, where $A$ and $J$ are the area and angular momentum of any axially symmetric closed stable minimal surface in an axially symmetric maximal initial data.

\ack
We would like to thank S.~Dain and P.~T.~Chru\'{s}ciel  for illuminating discussions and Robert Thompson for commenting on
the manuscript.



\end{document}